\documentclass[%
 reprint,
 amsmath,amssymb,
 aps,
]{revtex4-2}

\usepackage{graphicx}% Include figure files
\usepackage{dcolumn}% Align table columns on decimal point
\usepackage{bm}% bold math
\usepackage{amsmath,amssymb,amsthm,easybmat,verbatim}
\usepackage{mathtools}

\usepackage{color}

\usepackage{physics}
\usepackage[caption=false]{subfig}
\usepackage{hyperref}
\usepackage[T1]{fontenc}

\allowdisplaybreaks

\begin{document}

\preprint{APS/123-QED}

\title{Detecting Errors in a Quantum Network with Pauli Checks}%error filtering no error mitigation% Force line breaks with \\
% \thanks{A footnote to the article title}%

\author{Alvin Gonzales$^{1}$}
\email{agonza@siu.edu}
\author{Daniel Dilley$^1$}
\author{Bikun Li$^2$}
\author{Liang Jiang$^2$}
\author{Zain H. Saleem$^1$}

\affiliation{$^1$Mathematics and Computer Science Division, Argonne National Laboratory, Lemont, IL, 60439, USA}

\affiliation{$^2$Pritzker School of Molecular Engineering, University of Chicago, Chicago, IL, 60637, USA}

% \collaboration{CLEO Collaboration}%\noaffiliation

\date{\today}% It is always \today, today,
             %  but any date may be explicitly specified

\begin{abstract}
We apply the quantum error detection scheme Pauli check sandwiching (PCS) to quantum networks by turning it into a distributed multiparty protocol. PCS provides protection on the targeted qubits and generally requires less resource overhead than standard quantum error correction and detection codes. We provide analytical equations for the final fidelity and postselection rate for different PCS checks. We also introduce a recursive version of PCS that generates a family of distance 2 quantum codes that are locally equivalent to Calderbank-Shor-Steane (CSS) codes. Our analytical results are benchmarked against the Bennet-Brassard-Popescu-Schumacher-Smolin-Wooters (BBPSSW) protocol in comparable scenarios. 
% In addition, we derive that PCS on Bell states forms a distance 3 code. 
We also perform simulations with noisy gates for entanglement swapping and attain fidelity improvements. Lastly, we discuss various setups and graph state properties of PCS. 
% Moreover, in entanglement swapping, since we know that the desired output state is an EPR pair, we can always estimate the average fidelity and determine if the scheme is improving the performance.

\end{abstract}

%\keywords{Suggested keywords}%Use showkeys class option if keyword
                              %display desired
\maketitle
\makeatother
\newtheorem{definition}{Definition}[section]
\newtheorem{assumption}{Assumption}[section]
\newtheorem{theorem}{Theorem}[section]
\newtheorem{lemma}{Lemma}[section]
\newtheorem{conjecture}{Conjecture}[section]
\newtheorem{property}{Property}[section]
\newtheorem{corollary}{Corollary}[section]

%

% \usepackage{biblatex}
% \addbibresource{dqva.bib}

%

% \newcommand{\Field}[0]{\symbb{F}}
% \DeclareMathOperator{\join}{join}
% \DeclareMathOperator{\sdwidth}{width}
% \DeclareMathOperator{\leaves}{lvs}
% \DeclareMathOperator{\gcut}{cut}
% \DeclareMathOperator{\emb}{emb}
% \DeclareMathOperator*{\argmin}{arg\ min}
% \DeclareMathOperator*{\argmax}{arg\ max}
% % \newcommand{\ket}[1]{\lvert #1 \rangle}

% \newcommand{\sminus}[0]{\scalebox{0.75}[1.0]{\( - \)}}
% \newcommand{\union}[0]{\cup}
% \newcommand{\intersect}[0]{\cap}
% \DeclareMathOperator*{\bigast}{\scalerel*{\ast}{\sum}}

\section{Introduction}
Quantum networks promise to revolutionize classical communication. For example, with a reliable quantum network we can perform quantum state teleportation,  superdense coding \cite{nielsen2011quantumCompAndQuantInfo}, and distributed quantum sensing \cite{Proctor_2018MultiParamestInNetworkedQuantSens}. A quantum network consisting of long distance Einstein-Podolsky-Rosen (EPR) pairs along with a classical network is considered to be the most viable framework to achieve reliable quantum communication \cite{Briegel_1998QuantRepTheRoleOfImperfLocalOpsInQuantComm}. Local EPR pairs are generated in repeaters and end to end entanglement is achieved through entanglement swapping \cite{Zukowski_1993EventReadDetectBellExpViaEntSwap}. 

Since entanglement cannot be generated from product states, using local operations and classical communications (LOCC) alone \cite{Chitambar_2014EverythingYouAlwaysWantedToKnowAboutLocc}, sharing entanglement requires sending part of locally generated entangled states into the network as flying qubits. This exposes EPR pairs to network noise consisting of qubit loss and state decoherence. A standard method for combating errors is to use an entanglement purification protocol (EPP) such as Bennet-Brassard-Popescu-Schumacher-Smolin-Wooters (BBPSSW) \cite{Bennett_1996bbpssw} and Deutsch-Ekert-Josza-Macchiavello-Popescu-Sanpera (DEJMPS) \cite{Deutsch_1996DEJMPS}, which use copies of lower fidelity EPR pairs to produce a smaller number of higher fidelity EPR pairs. However, distillation faces the challenges of high photon loss and the difficulty of performing CNOT gates. A similar approach is to use quantum error detection or error correction codes to protect memory qubits. However, meeting the requirements for fault tolerance is a demanding task \cite{Aharonov_2008FaultTolQuantCompWithConstErrRate, Kim_2023EvidForTheUtilOfQCBeforeFaultTol}.

In this paper, we modify the quantum computing error detection method Pauli Check Sandwiching (PCS) \cite{Debroy_2020ExtendFlagGadgetsForLowOverCircVerif, Gonzales_2023PCS} into a distributed protocol and apply it to quantum networks. PCS has been demonstrated to improve the fidelity of hardware experiments on quantum computers \cite{liu_2022classicalSimsAsQuantErrMitigViaCircCut, VenDenBerg_2023SingleShotErrMitigByCohPauliChecks, shaydulin2023evidenceOfScalAdvForQAOAOnClassIntracProb, li2024_Qutracer}. It works by verifying the transformations of elements of the Pauli group and is a single shot quantum error detection protocol (can succeed in a single execution). PCS is designed to protect only the targeted qubits \cite{Debroy_2020ExtendFlagGadgetsForLowOverCircVerif} and as a result generally requires less quantum overhead than standard quantum error correction or detection codes \cite{Gonzales_2023PCS, VenDenBerg_2023SingleShotErrMitigByCohPauliChecks}. Moreover, the structure of PCS allows us to add and remove checks while knowing the theoretical impact on the error channel. It is known that quantum error correction codes (QECC)s and entanglement purification protocols have a direct correspondence \cite{Aschauer_2005QuantCommInNoisyEnv, Dur_EntPurAndQEC}. Thus, PCS is an ideal candidate error detection scheme to apply to the task of protecting qubits in a quantum network. In the network setting, the ideal channel of the flying and memory qubits is the identity channel, which allows us to use any Pauli group element as the Pauli check in PCS. 

The novel contributions of our paper are as follows. In the context of protecting Bell states, we provide analytical results for the output fidelity and postselection rate for different PCS checks. We also provide a recursive PCS scheme whose CNOT and qubit cost only increase polynomially as a function of the number of recursions. The recursive PCS X\&Z scheme generates a family of distance 2 quantum codes whose stabilizer generators have a maximum weight (non identity terms) of 4. In comparison to BBPSSW, PCS outperforms in fidelity and postselection rate under comparable scenarios and under most of the input fidelity range.   

Next, we provide simulation results with noisy gates for differing setups, including entanglement swapping, and show fidelity improvements.  Finally, we discuss some possible setups for utilizing PCS. Since the checks in PCS can be be seen as encoding and decoding operations, PCS can be incorporated similarly as quantum error detection codes. Moreover, the checks can be performed in a manner that preserves the structure of graph states. Thus, PCS is relatively easy to integrate into lossy schemes such as the all photonic quantum repeaters \cite{Azuma_2015AllPhotonicQuantRep, Hasegawa_2019ExpTimReversedAdaptiveBellMeasTowAllPhotonicQuantRep} or repeaters with quantum memories.

\section{Background}
\subsection{Pauli Check Sandwiching}
The $n$ qubit Pauli group is 
\begin{align}
    \mathcal{P}_n=\{I, X, Y, Z\}^{\otimes n}\times\{\pm 1, \pm i\}.
\end{align}
In Pauli Check Sandwiching (PCS) we use symmetries of the payload unitary $U$ to detect errors \cite{Gonzales_2023PCS}. Let $L_i,R_i\in \mathcal{P}_n$ such that
\begin{align}\label{eq:pcs_cond}
    L_iUR_i=U.
\end{align}
In the one layer PCS scheme we introduce an ancilla and sandwich $U$ with Pauli checks such that
\begin{align}&(\op{+}\otimes I +\op{-}\otimes L_i)U(\op{+}\otimes I +\op{-}\otimes R_i)\\
&=(\op{+}\otimes U +\op{-}\otimes L_iUR_i),
\end{align}
where the control is on the ancilla register and the targets are the registers that $U$ acts on. We refer to
\begin{align}
    (\op{+}\otimes I +\op{-}\otimes L_i)
\end{align}
as the left check and 
\begin{align}
    (\op{+}\otimes I +\op{-}\otimes R_i)
\end{align}
as the right check. The protocol is executed by measuring the ancilla in the Pauli Z basis and postselecting on the zero outcome. This protocol extends trivially to multiple checks by using more ancillas and controlling the pairs of checks on different ancillas. PCS can succeed in a single shot, which is necessary for quantum networks. Moreover, in the theoretical limit of noiseless checks, there exists checks (including non Pauli for general circuits) such that the postselected state is noiseless \cite{Gonzales_2023PCS}. In contrast to QECCs, PCS provides protection on the targeted qubits and generally incurs a lower resource overhead.

\subsection{Entanglement Swapping and Quantum Networks}
To create entanglement links between long distances, the standard protocol is to perform entanglement swapping. Quantum networks focus on sharing Bell states, which are all locally equivalent to the EPR pair
\begin{align}
    \ket{\Phi^+}=\frac{1}{\sqrt{2}}(\ket{00}+\ket{11}).    
\end{align}
Consider EPR pairs $\ket{\Phi^+}_{a_0a_1}$ and $\ket{\Phi^+}_{b_0b_1}$. Performing a Bell basis measurement on registers $a_0$ and $b_0$ results in $a_1$ and $b_1$ being in a Bell state. Entanglement swapping is basically the original quantum teleportation protocol \cite{Bennett_1993Teleportation}, but with the state teleported being one half of an EPR pair.

A standard figure of merit we will use is the fidelity
\begin{align}
    F(\rho_1,\rho_2)=\tr(\sqrt{\sqrt{\rho_1}\rho_2\sqrt{\rho_1}})^2,
\end{align}
between states $\rho_1$ and $\rho_2$. Throughout the text, the fidelity of a state $\rho$ is taken with respect to $\ket{\Phi^+}$ unless stated otherwise. This is given by
\begin{align}
    F(\rho, \ket{\Phi^+})=\tr(\rho\op{\Phi^+}).
\end{align} 

Since we know the desired output state, we can estimate the average fidelity of a generated Bell state by measuring the observables $XX$, $YY$, and $ZZ$ over an ensemble of generated Bell states because $\op{\Phi^+}=\frac{1}{4}(II+XX-YY+ZZ)$. Thus, it is always possible to determine if the the error detection/correction protocol is improving the fidelity of the output state.

Creating long links requires multiple executions of entanglement swapping. Since this is typically a non deterministic process, quantum repeaters are required to have quantum memories or use an all photonic setup like in \cite{Azuma_2015AllPhotonicQuantRep}.
% This is depicted in Fig.~\ref{fig:ent_swapping}.
% \onecolumngrid
% \begin{figure}[h!]
%     \centering
%     \includegraphics[width=\textwidth]{heralded_ent_fig.png}
%     \caption{Standard entanglement swapping. protocol}
%     \label{fig:ent_swapping}
% \end{figure}
% \twocolumngrid

\subsection{Entanglement Purification}
Recurrent entanglement purification methods use copies of an initial entangled state $\rho$ to produce a high fidelity EPR state. The BBPSSW protocol introduced by Bennett et al. \cite{Bennett_1996bbpssw}, can generate a higher fidelity state starting with two bipartite entangled states with initial fidelity $F>\frac{1}{2}$. BBPSSW relies on Werner states. The output fidelity given two initial copies of entangled pairs with fidelity F is
\begin{align}
    F'=\dfrac{F^2+\left[\frac{(1-F)}{3}\right]^2}{F^2+\frac{2F(1-F)}{3}+5\left[\frac{(1-F)}{3}\right]^2}
\end{align}
and the success probability is
\begin{align}
    c=F^2+\frac{2F(1-F)}{3}+5\left[\frac{(1-F)}{3}\right]^2.
\end{align}
Provided two copies ($\rho_{a_1b_1}$ and $\rho_{a_2b_2}$) of an entangled mixed state, BBPSSW proceeds by (i) depolarize each copy to Werner form (ii) apply bilateral CNOT gates (CNOT$_{a_1a_2}$ and CNOT$_{b_1b_2}$) (iii) measure one pair in the Pauli Z basis (iv) postselect the other pair if the measurement outcomes coincide. BBPSSW can be applied recursively with each round of the protocol requiring two CNOT gates and two copies of the output state of the previous round.

Error suppression can be improved by leveraging coordination between the recursion levels. In Ref.~\cite{Jiang_2007DistQuantCompBasedOnSmallQuantRegs}, a concatenated 2-way EPP was introduced that performs bit flip error suppression in the first level (using Z basis measurements) and phase flip error suppression (using X basis measurements) in the second level.

The DEJMPS protocol \cite{Deutsch_1996DEJMPS} is closely related to BBPSSW except DEJMPS uses initial states that have a Bell diagonal form. For one round, the two protocols share the same final fidelity and success rate. BBPSSW and DEJMPS are very efficient and in a specific scenario can be optimal \cite{Rozpedec_2018optimizingPracEntDistill}. DEJMPS and BBPSSW are 2-way entanglement purification protocols. Another class of purification methods are 1-way EPPs \cite{Azuma_2023QuantRepsReview}. An example of a 1-way EPP is hashing \cite{Bennett_1996MixedStEntAndQEC_hashing, Azuma_2023QuantRepsReview}, which relies on operating on a large ensemble of initial states. Hashing methods suffer greatly from gate errors \cite{Dur_EntPurAndQEC}.

% \subsection{Graph States}
% A graph state is a quantum state that can be represented by a graph and is a commonly used resource in quantum computing and quantum networks \cite{Raussendord_2001AOneWayQuantComp, Hein_2006EntInGraphStsAndItsApps, Meignant_2019DistGraphStsOverArbitraryQuantNets}. The edges in a graph state represent controlled-Z (CZ) gate operations and the nodes are initialized in the
% \begin{align}
%     \ket{+}=\dfrac{1}{\sqrt{2}}(\ket{0}+\ket{1})
% \end{align}
% state.

% Let $G(V,E)$ be a graph, where $V$ is the set of vertices and $E$ is the set of edges. The corresponding graph state for $G$ is \cite{Hein_2006EntInGraphStsAndItsApps}
% \begin{align}
%     \prod_{(a, b) \in E}\text{CZ}_{a,b}\ket{+}^{\otimes \abs{V}}.
% \end{align}

% Another way to specify graph states is to use its stabilizer group \cite{gottesman_1997stabilizerCodes}. Let $N_a$ be the neighbors of $a$ in the graph $G$. Then the generators of the stabilizer group of $G$ consist of the operators \cite{Hein_2006EntInGraphStsAndItsApps}
% \begin{align}
%     S_a=X_a\bigotimes_{k\in N_a} Z_k\quad\forall a\in G.
% \end{align}

% Some basic graph operations that we will use are edge deletion/addition and vertex deletion. To delete or add an edge to a graph state between vertices $a$ and $b$, we perform $\text{CZ}_{a,b}$ or $\text{CZ}_{b,a}$. To delete vertex $a$ from a graph state, we perform a Z-basis measurement on vertex $a$. The actions of single Pauli measurements on graph states are provided in Appendix \ref{appendix:local_corrections}.

\section{Results}

\subsection{Network PCS}
We extend the PCS protocol to a distributed protocol and explore its utility in quantum networks. Similar to quantum error correction and detection, PCS in general does not rely on destructive measurements of the data qubits. Thus, it can succeed in the single shot scenario and can be readily applied in the network setting.

In the setting of a quantum network, in Eq.~\eqref{eq:pcs_cond} $U$ is the identity channel. Therefore, $L_i=R_i$. The main idea in applying PCS is to perform the PCS protocol in a distributed fashion for flying qubits. For memory qubits, the operations are local. We describe here the distributed version. The network PCS protocol can be summarized in three steps: (1) implement the left checks at the origin repeaters on the qubits that will be sent, (2) send the flying qubits along with their ancillas, and (3) at the receiving repeater perform the right checks and postselect according to the standard PCS protocol.

For our numerical simulations with noisy gates, one scenario we examine is entanglement swapping with PCS. For this scenario, we consider the simple setting consisting of three repeater nodes: Alice, Bob, and Charlie arranged in a linear topology A-C-B. PCS can be executed as follows: (1) Alice and Bob prepare EPR states along with the left Pauli checks and send over two qubits each, (one data qubit and one anscila) to Charlie via an optical fiber link. These form the left end of the PCS scheme. Charlie then applies the right PCS checks along with entanglement swapping then does postselection by performing measurements on the ancilla qubits. Charlie announces the results to Alice and Bob.

% This is depicted in Fig.\ref{fig:npcs_overview}.
% \onecolumngrid
% \begin{widetext}    
% \begin{figure}
%     \centering
%     \includegraphics[width=0.8\textwidth]{npcs_overview.png}
%     \caption{NPCS left parts}
%     \label{fig:npcs_overview}
% \end{figure}
% \end{widetext}
% \twocolumngrid

\begin{figure}
    \centering
    
    \subfloat[\label{fig:pcs_x_circ} Entanglement swapping protocol with PCS X checks]{\includegraphics[width=0.48\textwidth]{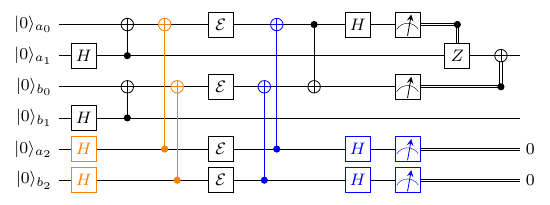}}
    
    \subfloat[\label{fig:npcs_xz} Entanglement swapping protocol with PCS X\&Z checks]{\includegraphics[width=0.48\textwidth]{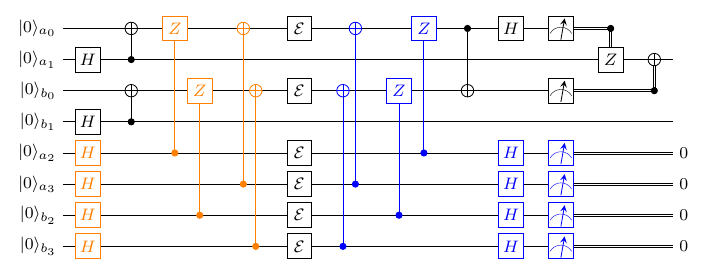}}
    \caption{Entanglement Swapping Protocol with PCS.  The orange gates make up the left Pauli checks and the blue gates make up the right Pauli checks. $\mathcal{E}$ denotes traversals through the network to Charlie and is a noise channel. We denote the origin repeater of a qubit by the subscript $a$ or $b$. Fig.~\ref{fig:pcs_x_circ} only has X checks, whereas Fig.~\ref{fig:npcs_xz} has both X and Z checks.}
    % \label{fig:npcs}
\end{figure}

The entanglement swapping protocol with PCS on flying qubits can be represented by the circuit in Fig. \ref{fig:pcs_x_circ}. We denote the origin repeater of a qubit by the subscript $a$ or $b$. At the end of the circuit and in the noiseless case, $a_1$ and $b_1$ is a Bell state. We can also include CZ checks by introducing more ancillas as shown in Fig.~\ref{fig:npcs_xz}. 
% Finally, note that from the nature of PCS, the structure of the entangled input bipartite state is not important and plays no role in the protocol.

% For the second case, repeater Charlie has two options, it can either post select on the zero outcomes of the ancilla qubits or use the information to try to perform corrections on the data qubits. 

\subsection{Theoretical Performance}
\begin{figure}
    \centering
    \subfloat[\label{fig:purWithBbpssw}Purification with BBPSSW]{\includegraphics[width=0.48\textwidth]{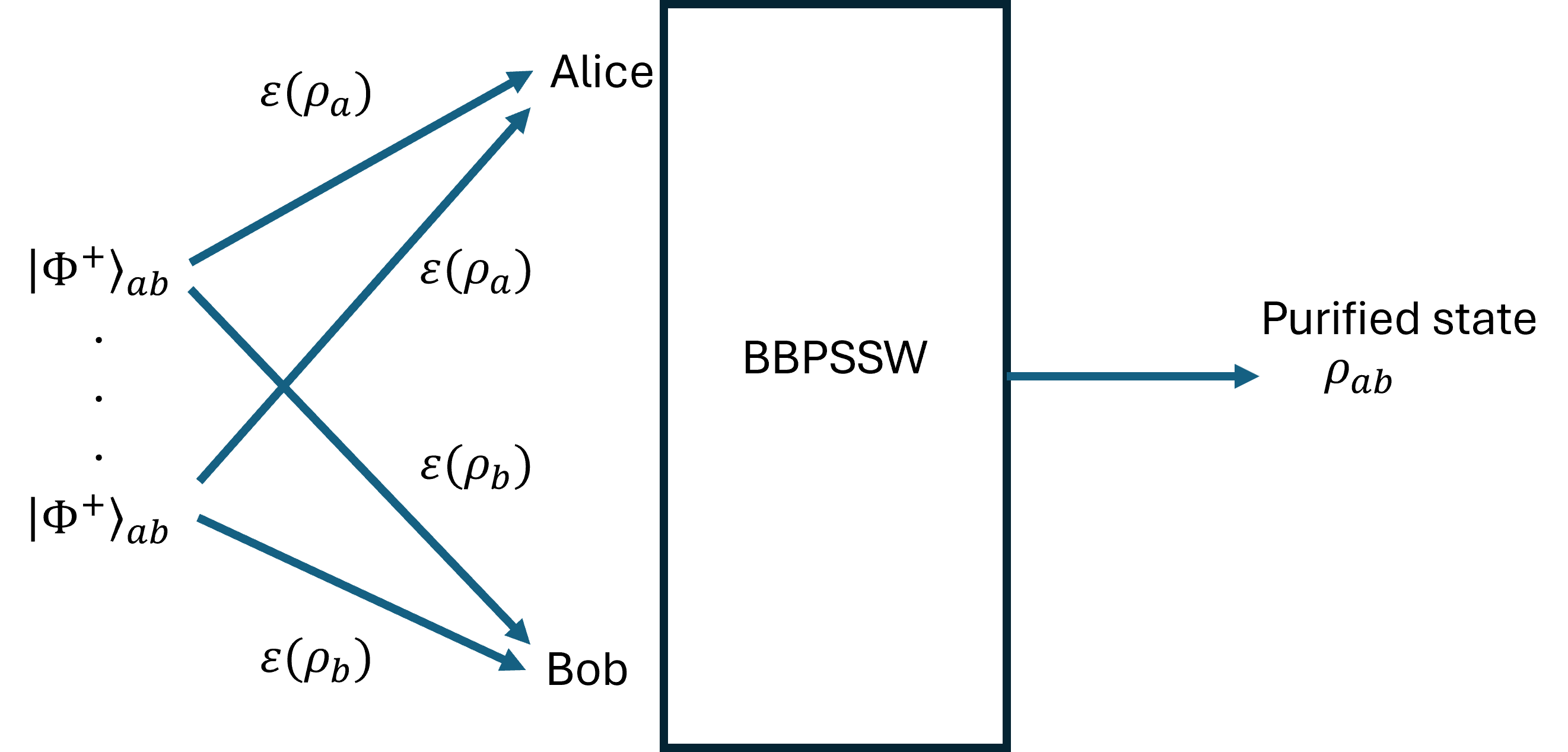}}
    
    \subfloat[\label{fig:purWithPCS}Purification with PCS]{\includegraphics[width=0.48\textwidth]{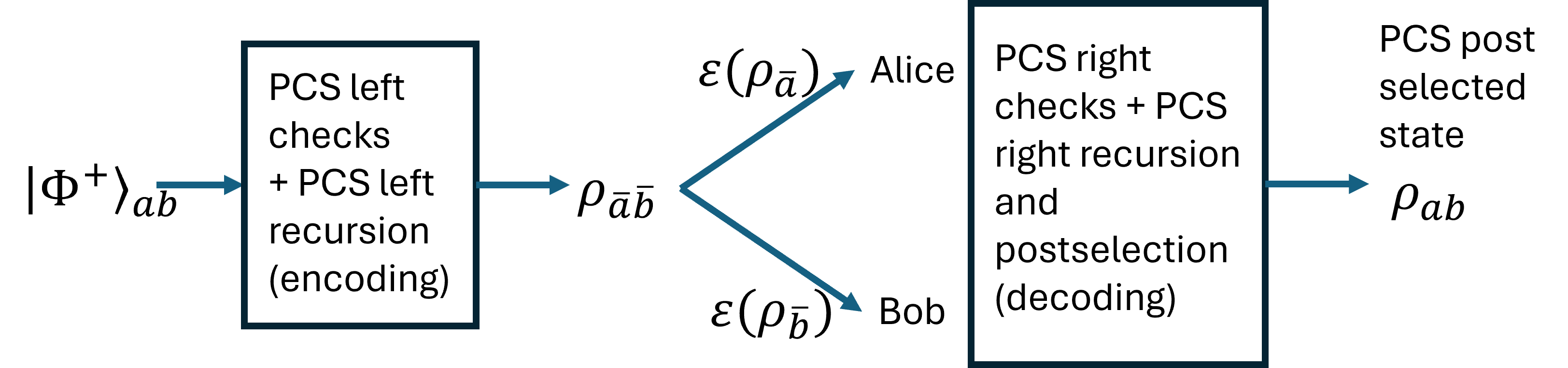}}
    \caption{PCS and BBPSSW purification scenarios. Note that having local noise channels on both halves of the Bell state in the BBPSSW protocol does not change as a function of the initial fidelity the original BBPSSW fidelity and postselection equations. The overline in the subscripts in the PCS scenario denotes multiple qubits. All the noise channels are single qubit depolarizing channels. \cite{Bennett_1996bbpssw}.}
    \label{fig:sketch_bbpssw_pcs}
\end{figure}

PCS offers flexibility in the type of checks used. In this section, we establish the base theoretical performance of PCS for a variety of reasonable setups. Note that we analyze the setting when there are only two parties Alice and Bob and no swapping is performed. This is the scenario considered by Bennett et al. in Ref.~\cite{Bennett_1996bbpssw}. The relationship between the two scenarios is described in Fig.~\ref{fig:sketch_bbpssw_pcs}. The checks in PCS detect anti commuting terms of the Kraus operators in the error channel \cite{Gonzales_2023PCS}. Thus, we consider the scenario consisting only of X checks and another scenario with X\&Z checks. We also investigate adding checks on the ancillas, which results in a family of distance 2 codes. Finally, we discuss using the ``ricochet'' property of Bell states with teleported PCS.

The theoretical performance of the PCS scheme can be expressed in terms of fidelity and postselection rate like in 2-way EPP protocols. A single qubit depolarizing channel can be defined as
\begin{align}
    \mathcal{E}(\rho)=\sum_{E_i}E_i\rho E_i^\dagger,
\end{align}
where $p$ is the probability of an error, $E_1=\sqrt{1-\dfrac{3p}{4}}I$ and $E_2,E_3,$ and $E_4$ are $\sqrt{\dfrac{p}{4}}X$, $\sqrt{\dfrac{p}{4}}Y$, and $\sqrt{\dfrac{p}{4}}Z$, respectively. Note that this is a completely depolarizing channel when $p=1$.

\subsubsection{PCS X Checks}
\begin{figure}
    \centering
    \includegraphics[width=0.48\textwidth]{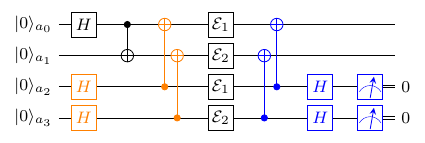}
    \caption{Analytical scenario for only PCS X checks.}
    \label{fig:pcs_x_fidelities_circ}
\end{figure}

Let \textit{all} the qubits undergo single qubit depolarizing channels, where $a_0$ and $a_2$ evolve under $\mathcal{E}_1(\rho)$ and $a_1$ and $a_3$ evolve under $\mathcal{E}_2(\rho)$ as shown in Fig.~\ref{fig:pcs_x_fidelities_circ}. $p_1$ and $p_2$ are the error probabilities of the depolarizing channels of $\mathcal{E}_1$ and $\mathcal{E}_2$, respectively.
Then the postselection rate is (see Appendix~\ref{appendix:PCS fidelity_postselect} for the derivation and final error Kraus map.)
\begin{align}
    c=\frac{1}{4} [(p_1-2) p_1+2] [(p_2-2) p_2+2]
\end{align}
and the final fidelity is
% \begin{widetext}
\begin{align}
    F'=\frac{\splitfrac{[9 (p_1-2) p_1+10] p_2^2+2 (20-9 p_1) p_1 p_2}{+2 p_1 (5 p_1-12)-24 p_2+16}}{4 [(p_1-2) p_1+2] [(p_2-2)
   p_2+2]}.
\end{align}
% \end{widetext}
These can be restated as 
\begin{align}
    c=\frac{1}{9}(1+2F)^2
\end{align}
and
\begin{align}
    F'=\frac{9F^2}{(1+2F)^2},
\end{align}
where $F$ is the initial fidelity (fidelity of the noisy Bell state without PCS) and we let $p_1=p_2=p$. The PCS X scheme requires 2 ancillas for a total of 4 physical qubits. A plot of the final fidelity vs the initial fidelity is shown in Fig.~\ref{fig:pcs_x_fidelities}. We compare it with the 1 round BBPSSW protocol which uses 4 qubits. The diagonal line represents the scenario where the input Bell state has the same fidelity as the output Bell state. A key point to note is that, a single PCS check can be used (1 ancilla) and it is possible to improve the fidelity of the initial Bell state as calculated in Appendix \ref{appendix:PCS fidelity_postselect}. This differs from many existing EPP, quantum error correction, and quantum error detection protocols.

\subsubsection{PCS X and Z Checks}
\begin{figure}
    \centering
    \includegraphics[width=0.45\textwidth]{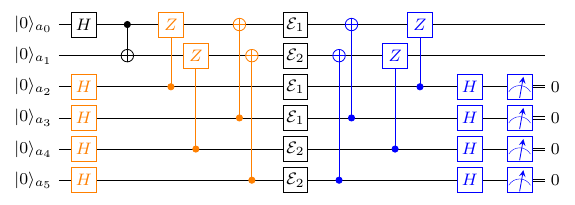}
    \caption{Analytical scenario for PCS X\&Z checks.}
    \label{fig:pcs_xz_fidelities_circ}
\end{figure}

For a Bell state protected by PCS X\&Z checks on both qubits (total of 6 physical qubits) as shown in Fig.~\ref{fig:pcs_xz_fidelities_circ}, we have a postselection rate of
\begin{align}
    c=\frac{1}{16} (p_1-2) [p_1 (2 p_1-3)+2] (p_2-2) [p_2 (2 p_2-3)+2]
\end{align}
and fidelity with the Bell state of
\begin{widetext}
\begin{align}
    F'=\frac{[p_1 (13 p_1-25)+14] p_2^2-25 (p_1-2) p_1 p_2+14 (p_1-2) p_1-28 p_2+16}{4 [p_1 (2 p_1-3)+2] [p_2 (2
   p_2-3)+2]}. 
\end{align}
\end{widetext}
These can be rewritten as
\begin{align}
    c=\frac{1}{324}(3+6F-\sqrt{12F-3}+4F\sqrt{12F-3})^2
\end{align}
and
\begin{align}\label{eq:pcsxz_fid}
    F'=\frac{1+52F^2-\sqrt{12F-3}-2F(4+\sqrt{12F-3})}{(\sqrt{12F-3}-1-8F)^2},
\end{align}
where we let $p_1=p_2=p.$
\begin{figure*}
    \centering
    \subfloat[We use PCS X checks on both qubits. There are 2 ancillas for a total of 4 physical qubits. In the PCS analysis, all the qubits experience the same depolarization. The 1 round BBPSSW uses 4 qubits.\label{fig:pcs_x_fidelities}]{\includegraphics[width=0.40\textwidth]{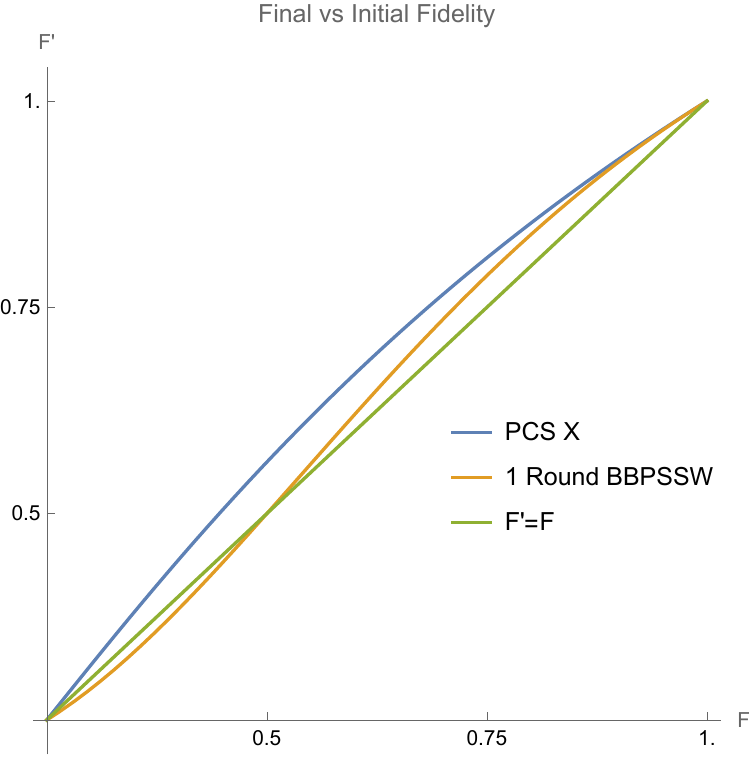}}\hfil
    \subfloat[We use PCS X\&Z checks on both qubits. The PCS method requires 4 ancillas for a total of 6 physical qubits. In the PCS analysis, all the qubits experience the same depolarization. The 2 round BBPSSW uses 8 qubits and 3 round BBPSW uses 16 qubits. \label{fig:pcs_xz_fidelities}]{\includegraphics[width=0.40\textwidth]{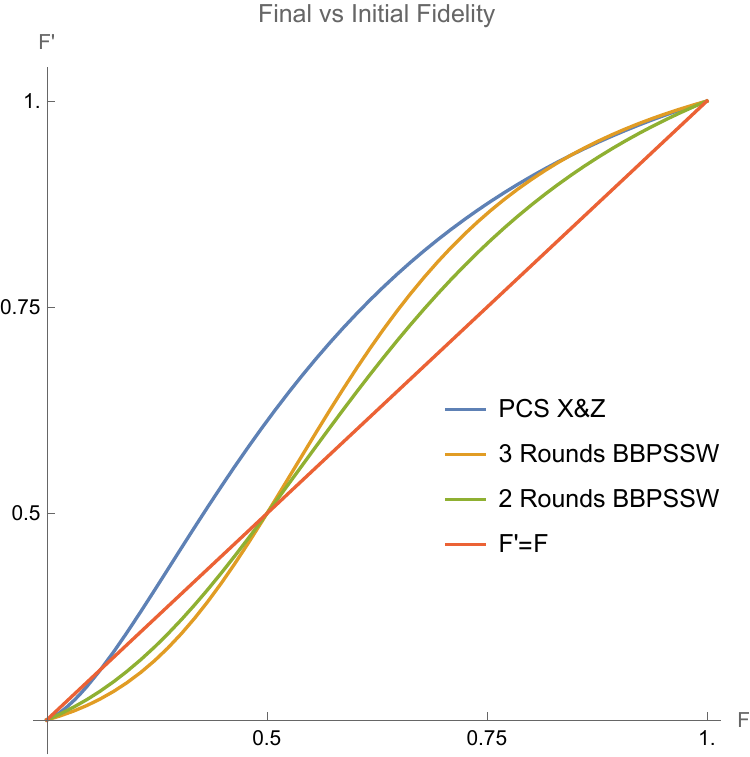}}
    \caption{Plots of the final fidelity $F'$ vs the initial fidelity $F$ of the noisy Bell state.}
\end{figure*}

The first order expansion of Eq.~\eqref{eq:pcsxz_fid} as a function of the infidelity $E=1-F$ is $1-E/3$. The PCS X\&Z scheme requires 4 ancillas for a total of 6 physical qubits. A plot of the fidelity as a function of the initial fidelity is given in Fig.~\ref{fig:pcs_xz_fidelities}. To benchmark the efficacy of PCS, we plot results from PCS and different recursion levels of BBPSSW. Note that BBPSSW is not the best multi round protocol \cite{Jiang_2007DistQuantCompBasedOnSmallQuantRegs}, since we can alternate between X and Z basis measurements for each round to achieve a higher fidelity. However, PCS X\&Z only utilizes 6 qubits, which is not enough for the concatenated protocol \cite{Jiang_2007DistQuantCompBasedOnSmallQuantRegs} which requires at least 8 qubits to perform concatenated purification. We compare PCS X\&Z with 2 rounds of the BBPSSW protocol which uses 8 qubits and 3 rounds of BBPSSW which uses 16 qubits. In the cases investigated, we outperform BBPSSW in terms of fidelity for most of the domain, while requiring less qubits.

\begin{figure}
    \centering
    \includegraphics[width=0.4\textwidth]{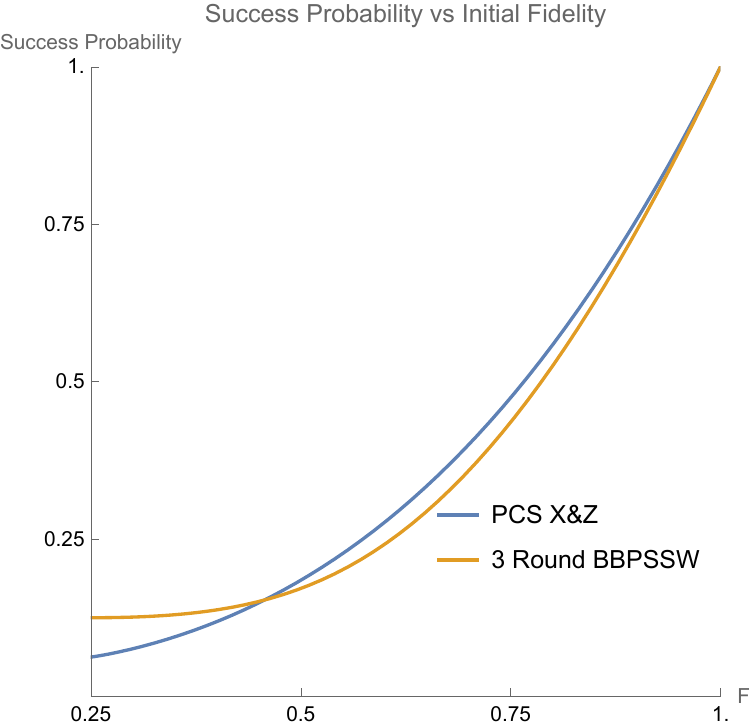}
    \caption{Success probability comparison. Note that we use 3 round BBPSSW because it is the minimum number of rounds that achieves comparable output fidelities to PCS over some ranges of input fidelity.}
    \label{fig:pcs_xz_success_probs}
\end{figure}

We also compare the success probabilities of PCS X\&Z with BBPSSW. We use 3 rounds of BBPSSW for comparison because that achieves, for initial high fidelities, a comparable output fidelity to PCS. As shown in Fig.~\ref{fig:pcs_xz_success_probs}, the two schemes exhibit similar success probabilities with PCS being slightly higher over most of the domain.

A subtle, but key point here is that we cannot recursively  apply the PCS equations because the ancilla error rates are also determined by the input fidelity $F$. Additionally, from the point of view of the errors detected by PCS \cite{Gonzales_2023PCS}, repeating the same checks with the same target qubits, but a different ancilla should offer little benefit.  This differs from EPP and error correction and detection, where you can recursively apply the technique. 

\subsubsection{Recursive PCS Checks}
\begin{figure}
    \centering
    \includegraphics{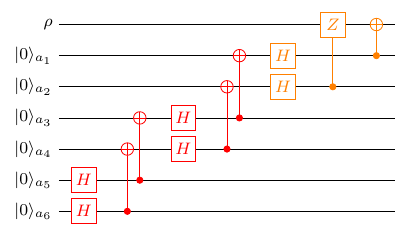}
    \caption{Recursive PCS X\&Z. Note that we only show part of the complete circuit (one of the Bell state qubits and its associated left checks). The recursive checks after recursion 1 protect the ancillas of the previous recursion level. The recursion gates are in red and there are 2 levels of recursion. Each recursion level uses 2 Hadamards, 2 CNOTS, and 2 qubits because we are only showing half of the Bell state.}
    \label{fig:pcs_circ_xz_recursion_circ}
\end{figure}

Instead, we design a recursion scheme where the recursion is performed by adding more PCS checks, but the targets of the check gates are ancillas from the previous recursion. This can be repeated as desired. The intuition behind this method is that the PCS X\&Z scheme postselects a perfect state if the PCS subcircuit is noiseless \cite{Gonzales_2023PCS}. Thus, the recursion is meant to improve the quality of the initial PCS checks. For each recursion level, we only apply PCS X checks on the ancillas of the previous level because Z errors on the ancillas are the errors that go undetected. Each recursion level introduces two additional X checks. A diagram of this construction, shown for the left checks part of the circuit, is provided in Fig.~\ref{fig:pcs_circ_xz_recursion_circ}, where the red gates are from 2 recursion levels and the orange gates are the base PCS X\&Z checks. 

% Note again that the targets of the next recursion of PCS check gates are the ancilla qubits of the previous recursion. 
Note that we do not provide the analytical recursive fidelity equation, but numerical results are provided in Fig.~\ref{fig:pcs_xz_recursion_plot}. 
% for the circuit given in Fig.~\ref{fig:pcs_xz_fidelities_circ} with additional recursive checks. 
A related scheme was introduced in Ref.~\cite{VenDenBerg_2023SingleShotErrMitigByCohPauliChecks} for quantum computing, but here we apply it in a distributed setting, use specific checks and a specific target state, investigate its relation to error correction, and we examine its scaling. 

% \subsubsection{PCS Recursion Scaling}
The number of qubits and CNOT gates required for recursive PCS X\&Z, are only polynomial functions of the number of recursion levels. From Fig.~\ref{fig:pcs_circ_xz_recursion_circ}, the cost in terms of the number of ancilla qubits and the number of CNOTs for the right checks are each given by the function 
\begin{align}
    C(r)=4(r+1),
\end{align}
where $r$ is the recursion level and the 4 is appears since the circuit shows only one of the parties.
% Given the results of our numerical simulations in Fig.~\ref{fig:pcs_xz_recursion_plot}, the yield of the recursive PCS scheme is very high. 
In practice we can vary the amount of protection each recursion provides. For instance, we can choose not to protect all the ancillas from the previous recursion or we can use different checks. We can do this because PCS is a flexible method. Thus, we can reduce the cost of recursion if desired.

\subsubsection{Relation to Quantum Error Correction}
The PCS scheme can be viewed from the perspective of quantum error correction with stabilizer codes. The left PCS gates perform encoding and the right PCS gates perform decoding. We can determine the stabilizers of the state after the left PCS X\&Z checks (orange gates) in Fig.~\ref{fig:pcs_xz_fidelities_circ}. PCS X\&Z can be considered a distance 1 code because it cannot detect an arbitrary single qubit error. However, PCS X\&Z can detect arbitrary single qubit errors on the targeted qubits. 

The recursive PCS X\&Z (shown with two recursions in Fig.~\ref{fig:pcs_circ_xz_recursion_circ}) generates a family of distance 2 codes \cite{gottesman_1997stabilizerCodes}. With 1 recursion the scheme forms a [[5, 1, 2]] code. With 2 recursions the scheme forms a [[7, 1, 2]] code. Thus, this scheme generates a [[2$(r-1)+5$, 1, 2]] code for recursion $r\geq 1$. An interesting property of these codes is that the weight (non identity components) of the generators of their stabilizer groups are upper bounded by 4 irrespective of the number of recursions. There is only one generator that has weight 4 and it is given by $Z_\rho Z_{a_1}X_{a_2}Z_{a_4}$. The family of codes are locally equivalent to Calderbank-Shor-Steane (CSS) codes through $H$ gates at appropriate locations after the encoding. The transformation to a CSS code follows a pattern, where we apply $H_{a_2}H_{a_3}$, $H_{a_6}H_{a_7}$, $H_{a_{10}}H_{a_{11}}$, etc. after the left checks of the recursive PCS X\&Z circuit.

Note that some single qubit errors can be uniquely identified and corrected by this encoding. Let us consider the complete PCS circuit for [[5, 1, 2]]. There is only one single qubit error that generates the error syndrome [0010], where we arranged the ancilla measurement outcomes as $a_1a_2\cdots a_4$. The unique single qubit error before the right checks that generates this syndrome is $Z_{a_3}$. However, some higher weight errors can generate this same syndrome such as $X_{\rho}X_{a_1}$ before the right checks. Thus, the tradeoff is an increase of the postselection rate by the first order of the error rate $p$, while  decreasing the fidelity by the second order of $p$. 

\subsection{Teleported PCS}
\begin{figure}
    \centering
    \includegraphics[width=0.48\textwidth]{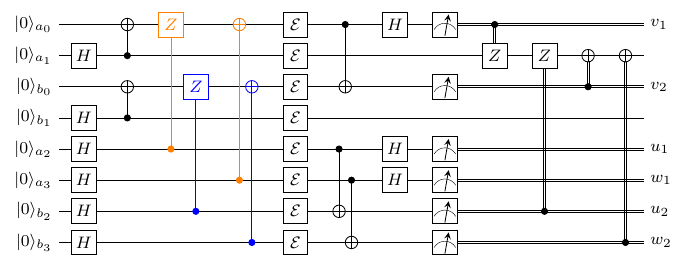}
    \caption{Teleported PCS X\&Z. In the scenario of entanglement swapping, we can perform a variant of the PCS scheme which utilizes additional Bell state measurements across the ancillas $a_2, b_2$ and $a_3, b_3$. The postselection criteria is $w_1+v_1=0$ and $u_1+v_2+w_2=0$. The effect is that less gate operations are required. For certain noise regimes, teleported PCS X\&Z outperforms standard PCS.}
    \label{fig:teleportedPCS}
\end{figure}

For the scenario of entanglement swapping, a variant of the PCS scheme can be constructed by applying Bell state measurements across the ancillas as shown in Fig.~\ref{fig:teleportedPCS}. In effect, the sandwiching of the noise is performed by the initial PCS encoding gates. The sandwiching effect occurs because the Bell state measurements connect qubits $a_0$ and $b_0$; $a_2$ and $b_2$; and $a_3$ and $b_3$. In the noiseless case only certain Bell state measurements can occur and thus, they can be used to construct the criteria for postselection. These possible outcomes can be extracted through the stabilizers of the noiseless state before the measurements. The postselection criteria for teleported PCS is $w_1+v_1=0$ and $u_1+v_2+w_2=0$.

% Since we are only interested in protecting the specific state $\ket{\Phi^+}$, we have 6 generators in total. The generators of the stabilizer group are
% \begin{align}
% &X_{a_5}X_{a_1}, X_{a_4}Z_{a_5}Z_{a_1}, X_{a_3}X_{a_0}, X_{a_2}Z_{a_3}Z_{a_0},\\
% \notag &Z_{a_1}Z_{a_5}Z_{a_3}Z_{a_0}, X_{a_0}X_{a_1}Z_{a_2}Z_{a_4}.
% \end{align}
% For any single qubit Pauli error, there exists at least one stabilizer that anticommutes with the error. Thus, any single  qubit error is detectable. For 2 qubit errors, notice that some $X_{a_i}X_{a_j}$ errors commute with all the stabilizers and are undetectable. However, these are elements of the stabilizer group. For 3 qubit errors, $Y_{a_5}X_{a_4}Y_{a_1}$ is an example of an undetectable error that is not a stabilizer of the state. Since we can detect up to 2 arbitrary single qubit errors, this is a distance 3 code. However, if the state is partitioned between two parties $(a_0,a_2,a_3)$ and $(a_1,a_4,a_5)$ and we are restricted to local operations, the stabilizers $Z_{a_1}Z_{a_5}Z_{a_3}Z_{a_0}, X_{a_0}X_{a_1}Z_{a_2}Z_{a_4}$ cannot be measured. In this restricted scenario, not all single qubit errors are detectable, e.g., $X_{a_2}$. It is an open question how we can use the information from PCS for correction without substantially degrading the fidelity.

\subsection{Simulations with Noisy Operations}
\begin{figure*}
    \centering
    \subfloat[\label{fig:pcsXZ_flyingq_equal_depol}]{\includegraphics[width=0.5\textwidth]{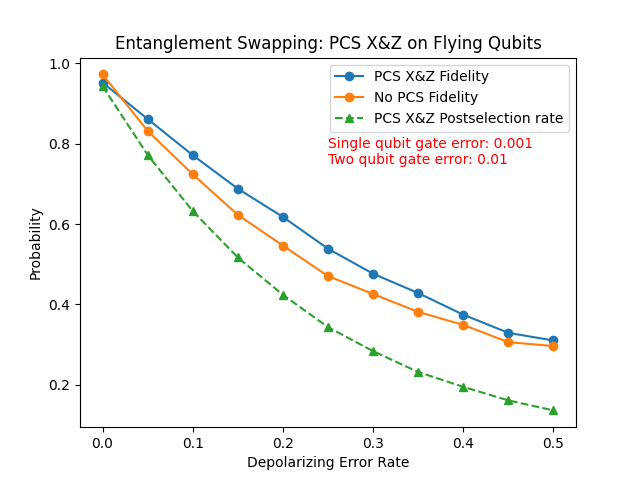}}
    \subfloat[\label{fig:pcsXZ_fqandmq_equal_depol}]{\includegraphics[width=0.5\textwidth]{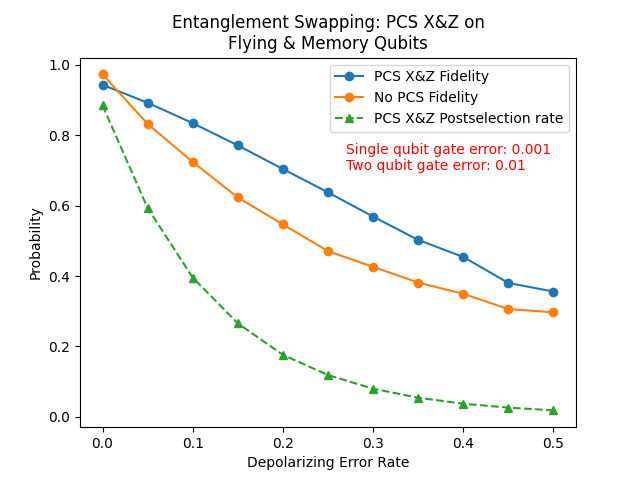}}
    \caption{Fig.~\ref{fig:pcsXZ_flyingq_equal_depol} is an entanglement swapping simulation with protection only on the flying qubits. The results are similar when only memory qubits are protected.  Fig.~\ref{fig:pcsXZ_fqandmq_equal_depol} is the simulation result with PCS X\&Z checks on both memory and flying qubits.}
    \label{fig:pcs_stabilizer_equiv}
\end{figure*}

\begin{figure}
    \centering
    \includegraphics[width=0.48\textwidth]{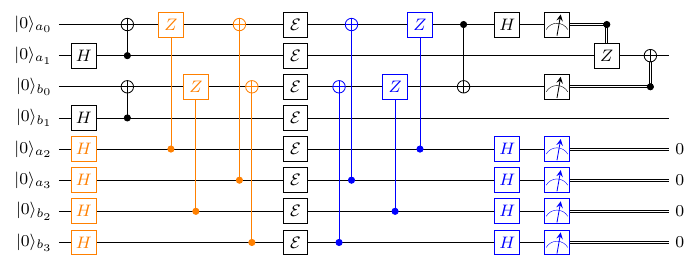}
    \caption{Entanglement swapping where PCS X\&Z checks are used to protect the flying qubits.}
    \label{fig:pcs_circ_xz_flying_all_noisy-label}
\end{figure}

In practice, gate operations are also noisy. In the following simulations, the single qubit gate and two qubit gate depolarization error rates are 0.001 and 0.01, respectively, which may be reachable in the future as hardware advances \cite{Shi_2022HighFidelityPhotonQuantLogGateBaseOnNearOptRydbergSinglePhotonSource}. The depolarizing error rate for the transmission channel $\mathcal{E}$ is varied from 0 to 0.5. First, we simulate the important case of entanglement swapping, which is a crucial protocol for extending the entanglement links in a quantum network.  The setup where flying qubits are protected by PCS X\&Z is shown in Fig.~\ref{fig:pcs_circ_xz_flying_all_noisy-label}.
Fig. \ref{fig:pcsXZ_flyingq_equal_depol} is the result when only flying qubits are protected and Fig.~\ref{fig:pcsXZ_fqandmq_equal_depol} is the result when we protect both flying and memory qubits. As expected, PCS on both the flying and memory qubits generally outperforms PCS on the flying qubits only, since the former provides more protection. However, this comes at a cost of a lower postselection rate. Note that the postselection rate at 0 depolarizing rate for the error channel is not 1, since the gate operations are noisy.

\begin{figure*}
    \centering
    \includegraphics[width=0.7\textwidth]{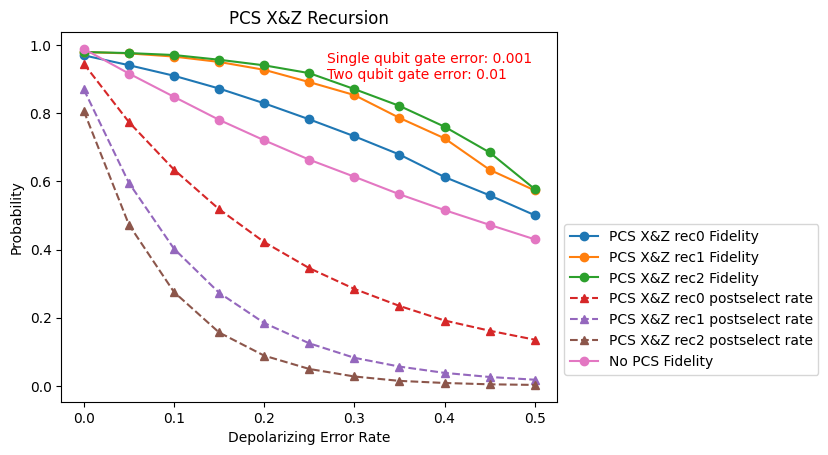}
    \caption{Simulation of recursive PCS X\&Z with noisy gates. Recursion 0 means the standard PCS X\&Z scheme. Recursion 1 and 2 means we protect all the initial ancillas with additional PCS X checks. The base PCS X\&Z scheme uses 6 qubits (4 ancillas and 2 from the Bell state). Recursion 1 uses 10 qubits and recursion 2 uses 14 qubits.}
    \label{fig:pcs_xz_recursion_plot}
\end{figure*}

Fig.~\ref{fig:pcs_xz_recursion_plot} is the numerical results when we apply recursion to the circuit in Fig.~\ref{fig:pcs_xz_fidelities_circ} (recursive PCS X\&Z). Note that 0 recursion means the standard PCS X\&Z scheme and 2 recursion means that we have performed the PCS recursion twice. As shown, the fidelity improvement is substantial. The base PCS X\&Z scheme uses 6 qubits (4 ancillas and 2 from the Bell state). Recursion 1 consists of 10 total qubits and recursion 2 consists of 14 total qubits. Since recursive PCS forms a family of distance 2-codes, adding more layers of recursion beyond rec1 does not provide much improvement.
\null\newpage
\subsection{Utilizing PCS in Repeaters}
% \begin{widetext}
% \onecolumngrid

% \twocolumngrid

% \end{widetext}
Since PCS can be seen as a distance 1 error detection code, we can utilize it in a similar manner as quantum error detection codes. We provide some examples for completeness. In repeaters with memory qubits, the memory qubits in the source repeaters and flying qubits are protected by PCS. The memory qubits of receiving repeaters are not protected. However, the decoherence experienced by these qubits can be considered small because in theory they can be initialized at the estimated arrival time of the photon. The receiver repeaters also do not have to wait for Bell state measurement (BSM) heralding signals from other repeaters. Still, if the idle time of the memory qubits of the receiver repeaters cause significant decoherence, we can also protect them before the entanglement with a photon occurs. 

\begin{figure*}
    \centering
    \subfloat[\label{fig:standard_mem_rep}]{\includegraphics[width=\textwidth]{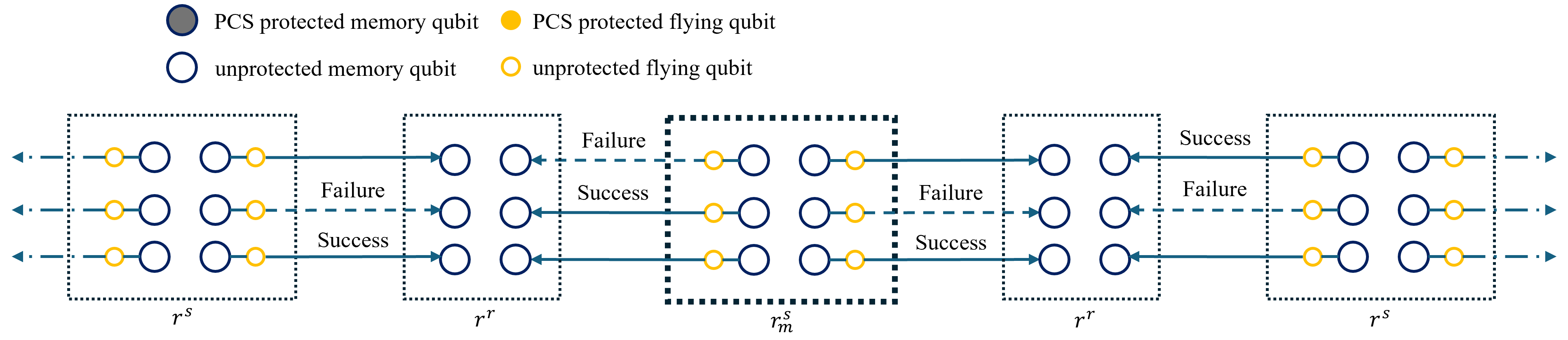}}
    
     \subfloat[\label{fig:pcs_rep1}]{\includegraphics[width=\textwidth]{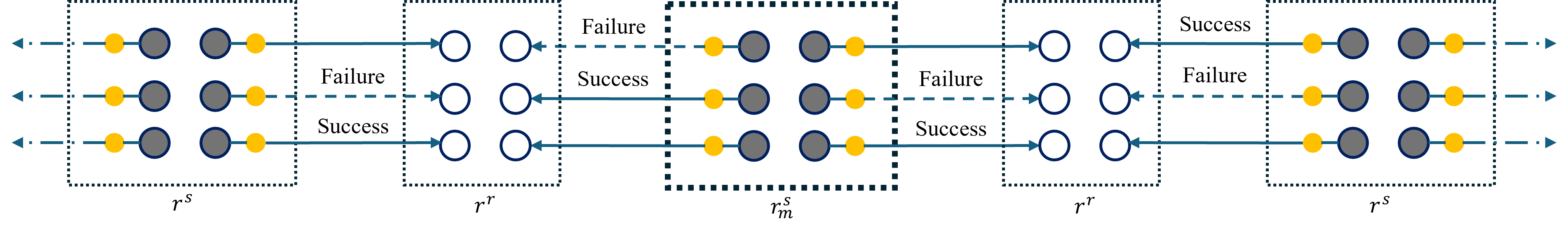}}
    
    \subfloat[\label{fig:pcs_rep2}]{\includegraphics[width=\textwidth]{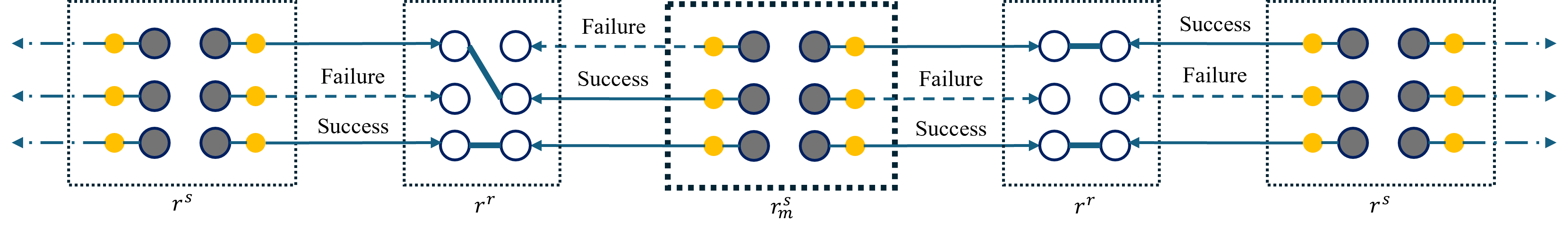}}

    \subfloat[\label{fig:pcs_rep3}]{\includegraphics[width=0.47\textwidth]{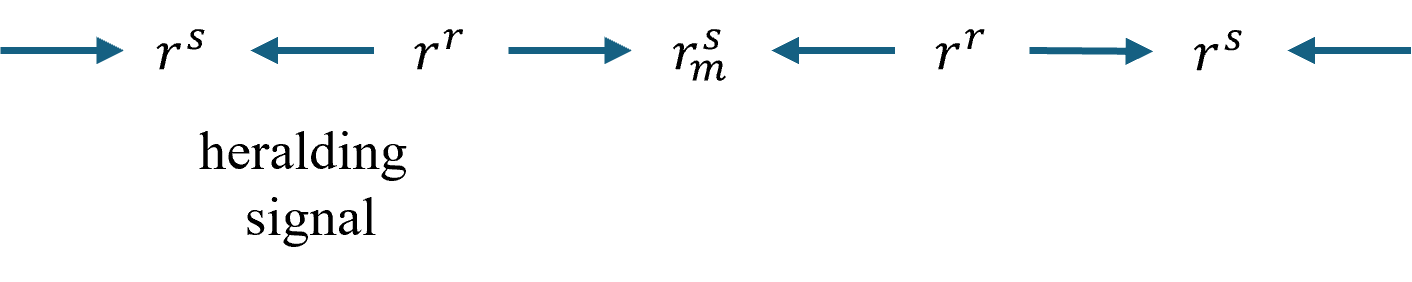}}
    
    \subfloat[\label{fig:pcs_rep4}]{\includegraphics[width=0.47\textwidth]{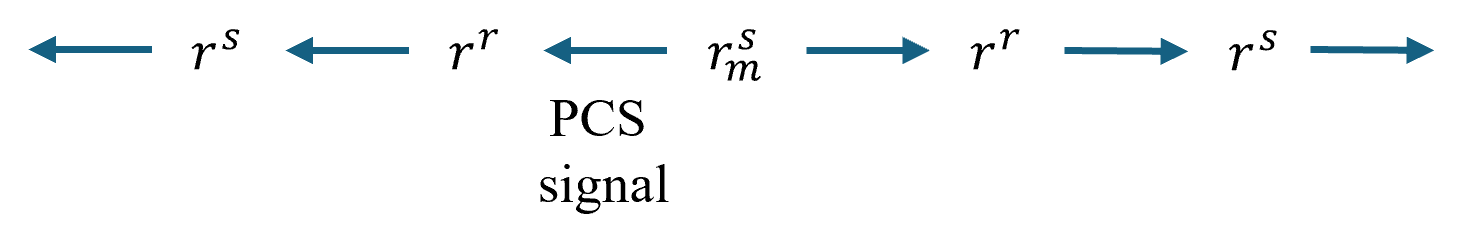}}
    \caption{Example method of incorporating PCS into known repeater setups. PCS can be seen as a distance 1 error detection code and thus can be utilized in a similar fashion. Fig.~\ref{fig:standard_mem_rep} consists of standard memory quantum repeaters without any purification, error detection, or error correction protocol. We can incorporate PCS by protecting the source repeaters' memory qubits and the flying qubits. The memory qubits of the receiver repeaters are unprotected, but the decoherence experienced by these unprotected memory qubits can be considered small because they are initialized at the estimated arrival time of the photons and they do not have to wait for heralding signals from other repeaters. In step 1, source repeaters send flying qubits to receiver repeaters and try to create entanglement with receiver memory qubits as shown in Fig~\ref{fig:pcs_rep1}. In step 2, the receiver repeaters perform local BSM measurements on the successful memory qubits as shown in Fig.~\ref{fig:pcs_rep2}. In the third step Fig.~\ref{fig:pcs_rep3}, the heralding signals are broadcasted from the repeater receiver nodes. Step 4 consists of entanglement swapping at the source repeaters. It begins when all the heralding signals have arrived. In the fourth step Fig.~\ref{fig:pcs_rep4}, the PCS postselections and local BSMs on the memory qubits are performed in a cascading fashion on the source repeaters starting at the middle source repeater. Except for the middle source repeater, which starts the process, a source repeater does not perform the PCS postselection and BSMs until it receives a PCS signal.}
    \label{fig:pcs_reps_proc_all}
\end{figure*}

After, an initial entanglement linking between a source and receiver repeater, the receiver repeater broadcasts its heralding signal(s). The source repeaters idle until they receive a heralding signal from each neighbor receiver repeater and except for the middle (or near the middle) source repeater, the source repeaters also wait for a PCS signal. Then the middle (or near the middle) source repeater completes its local PCS scheme and local entanglement swapping. The results are then sent to its neighboring source repeaters. This proceeds in an outward fashion until the end points. An example diagram of this process is given Fig.~\ref{fig:pcs_reps_proc_all}

% \begin{widetext}
% \onecolumngrid

% \twocolumngrid
% \end{widetext}

% \begin{figure}
%     \centering
%     \subfloat[\label{fig:pcs_rep3}]{\includegraphics[width=0.47\textwidth]{pcs_repeaters3.png}}
    
%     \subfloat[\label{fig:pcs_rep4}]{\includegraphics[width=0.47\textwidth]{pcs_repeaters4.png}}
    
%     \caption{In the third step Fig.~\ref{fig:pcs_rep2}, the heralding signals are broadcasted from the repeater receiver nodes. Step 4 begins when the all the heralding signals have arrived. In the fourth step Fig.~\ref{fig:pcs_rep3}, the PCS postselctions are performed in a cascading fashion starting at the middle source repeater. Except for the middle source repeater(s), which starts the process, a source repeater does not perform the postselection until it receives a PCS signal.}
%     \label{fig:pcs_repsb}
% \end{figure}

% \subsection{PCS on Graph States}
Another important repeater network design is the all photonic quantum repeaters \cite{Azuma_2015AllPhotonicQuantRep}. A key point in all photonic quantum repeaters is the utilization of graph states and time reversed adaptive Bell state measurements to create robustness against photon loss and remove idle time for the arrival of heralding signals from Bell state measurements \cite{Azuma_2015AllPhotonicQuantRep, Hasegawa_2019ExpTimReversedAdaptiveBellMeasTowAllPhotonicQuantRep}. A graph state is a quantum state that can be represented by a graph and is a commonly used resource in quantum computing and quantum networks \cite{Raussendord_2001AOneWayQuantComp, Hein_2006EntInGraphStsAndItsApps, Meignant_2019DistGraphStsOverArbitraryQuantNets}. As shown in Figs.~\ref{fig:pcs_rep2} and Fig.~\ref{fig:pcs_rep1}, the source repeaters have to wait for the heralding signals from the neighboring receiver repeaters to arrive before they can determine which memory qubits to connect. To avoid this waiting time, the source repeaters use highly entangled graph states and the receiving repeaters are memoryless \cite{Azuma_2015AllPhotonicQuantRep}. The repeaters can perform lossy Z measurements \cite{Azuma_2015AllPhotonicQuantRep, Varnava_2006LossTolInOneWayQuantCompViaCounterfactErrCorr} to disconnect lost qubits. 

\begin{figure}
    \centering
    \includegraphics[width=0.48\textwidth]{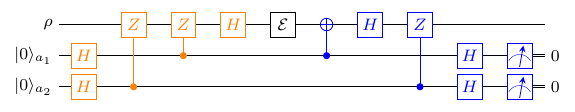}
    \caption{Example construction of lossy PCS X checks for graph states. Moving the orange $H$ gate on qubit wire $\rho$ after the error channel $\mathcal{E}$ converts these to lossy Z checks. $\rho$ is part of a larger graph state. Note that the state before $\mathcal{E}$ is locally equivalent to a graph state.}
    \label{fig:graph_pcsX&Z_construction}
\end{figure}

\begin{figure}
    \centering
    \includegraphics[width=0.4\textwidth]{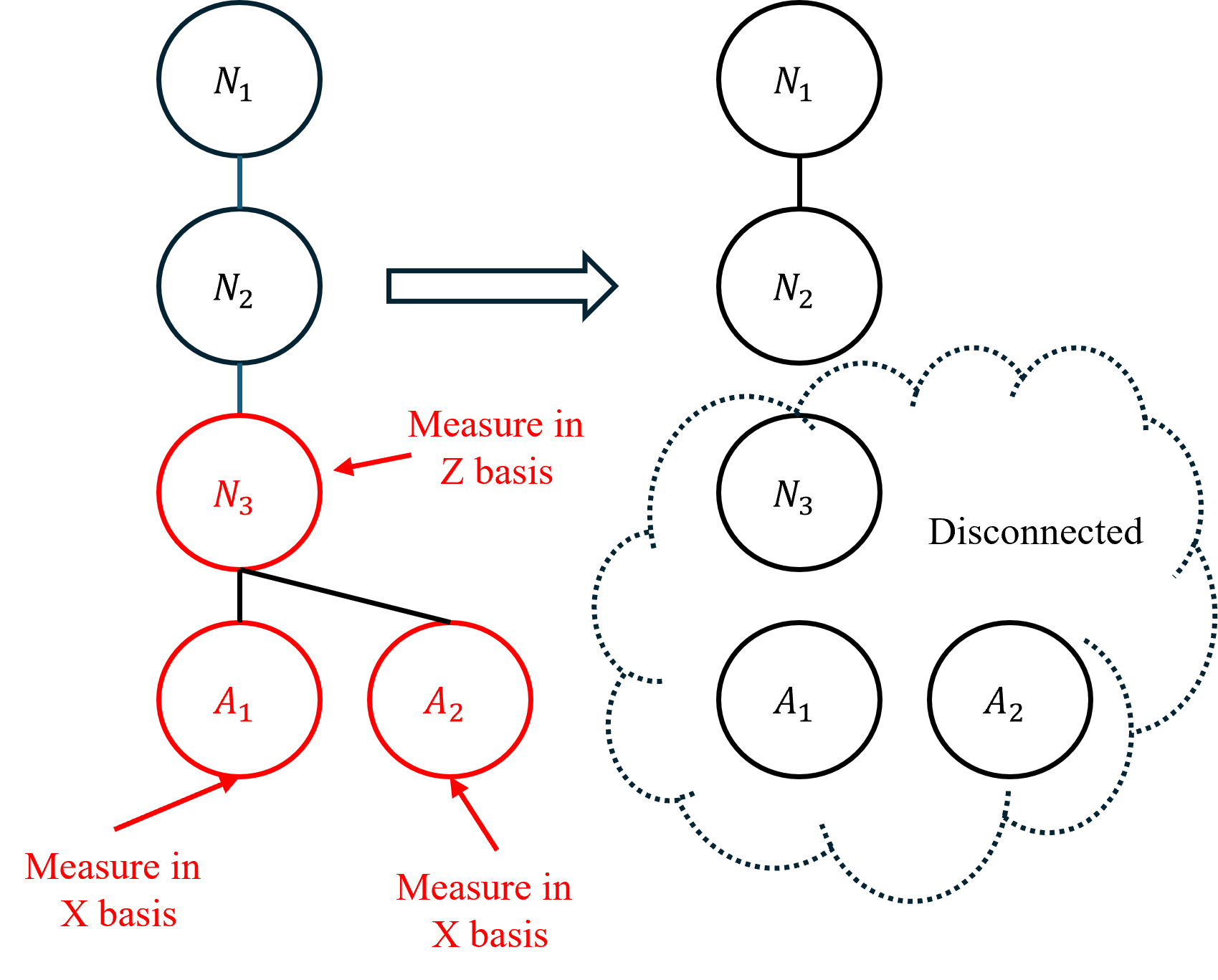}
    \caption{Consider an arbitrary graph state, where node $N_3$ has three neighbors. A Pauli Z basis measurement on $N_3$ or a Pauli X measurement on either ancilla qubits $A_1$ and $A_2$ disconnects the entire section from the rest of the graph state. Namely, any single operation in red results in the state to the right where the qubits in the cloud are in some form disconnected from the graph state $N_1$ and $N_2$ up to some local corrections. This has the structure of PCS, but with a possible rotated measurement on $N_3$ because of the PCS structure as shown in Fig.~\ref{fig:graph_pcsX&Z_construction}.}
    \label{fig:pcs_lossy}
\end{figure}

PCS can be naturally incorporated into the structure, by attaching the checks to the second leaf qubits \cite{Azuma_2015AllPhotonicQuantRep}. An example of how to perform lossy PCS X checks on an arbitrary graph state is shown Fig.~\ref{fig:graph_pcsX&Z_construction}. Additionally, PCS naturally fits into the tree construct of Varnava et. al.'s protocol \cite{Varnava_2006LossTolInOneWayQuantCompViaCounterfactErrCorr}. The tree structure allows us to disconnect a qubit as long as some of its descendants survive. An example is provided in Fig.~\ref{fig:pcs_lossy}. As long as one of the three qubits $N_3, A_1$, or $A_2$ survive, the entire three qubits can be disconnected by performing a lossy indirect Z basis measurement \cite{Varnava_2006LossTolInOneWayQuantCompViaCounterfactErrCorr}. If the ancillas are lost, we can measure the data qubit in the Z basis. If the data qubit is lost, we can measure either of the ancillas in the X basis. Majority voting can be incorporated in a similar fashion as in Varnava et al.'s protocol ~\cite{Varnava_2006LossTolInOneWayQuantCompViaCounterfactErrCorr}. Actions of local Pauli measurements on graph states are provided in Appendix \ref{appendix:local_corrections}. The state after PCS is locally equivalent to a graph state so we can use the same method as Varnava et. al.'s protocol \cite{Varnava_2006LossTolInOneWayQuantCompViaCounterfactErrCorr} except the measurement on $N_3$ is in the X basis for lossy PCS X checks.

\section{Conclusions}
We apply PCS to quantum networks and derive its analytical performance for various setups. PCS is a quantum computing error detection technique that provides protection on targeted qubits and thus, generally requires fewer gates and qubits to implement than standard error detection and correction codes. Since quantum codes and EPPs have a correspondence, PCS is a good candidate for use in networks. To benchmark its efficacy, we compared against BBPSSW. In terms of fidelity, it outperforms BBPSSW round 1, 2, and 3 for a comparable number of input qubits and over most of the values of the initial fidelity. It also outperforms BBPSSW in terms of postselection rate for similar scenarios and most of the values of the initial fidelity. We also introduce a recursive PCS scheme. Recursive PCS X\&Z generates a family of distance 2 codes that are locally equivalent to CSS codes and have a maximum stabilizer generator weight of 4.

% Next, we showed that PCS X\&Z on Bell states forms a distance 3 code. If or how this can be used for correction without degrading the fidelity is an open question. 

Our results are corroborated by numerical simulations with noisy gates. A potential theoretical line of research is to further investigate the recursive nature of PCS. Additionally, PCS ancillas follow a graph state like structure. For later generation repeaters, the graph properties of PCS could prove to be useful. 

\section{Code and Data Availability}
The code and data for numerical experiments are provided at \url{https://github.com/alvinquantum/quantum_networks_PCS}.

\section{Competing Interests}
The authors declare no competing financial or non-financial interests.

\section{Author Contributions}
AG formulated distributed PCS and recursive PCS,  derived the analytical equations, constructed a lossy PCS scheme, created figures, and performed experiments. DD provided useful Mathematica code, helped derive the analytical equations, and performed experiments. BL formulated teleported PCS, performed experiments, and determined the quantum error correction structure of recursive PCS. LJ and ZS supervised the research. All authors contributed to the writing of the manuscript.

\section{Acknowledgements}
AG thanks Joseph Chapman and Nicholas Peters from Oak Ridge National Laboratory for helpful discussions. This material is based upon work supported by the U.S. Department of Energy, Office Science, Advanced Scientific Computing Research (ASCR) program under contract number DE-AC02-06CH11357 as part of the InterQnet quantum networking project. B.L. and L.J. also acknowledge support from the ARO (W911NF-23-1-0077), ARO MURI (W911NF-21-1-0325), AFOSR MURI (FA9550-19-1-0399, FA9550-21-1-0209, FA9550-23-1-0338), DARPA (HR0011-24-9-0359, HR0011-24-9-0361), NSF (OMA-1936118, ERC-1941583, OMA-2137642, OSI-2326767, CCF-2312755), NTT Research, Packard Foundation (2020-71479).

\appendix
\section{Local Pauli Measurement on Graph States}\label{appendix:local_corrections}
We provide the effects of local Pauli measurements on graph states \cite{Hein_2006EntInGraphStsAndItsApps}. We measure a node $a$ that is a part of a larger graph state. Let $N_i$ define the neighbors of node $i$ and $\tau_i$ be the local complementation of the graph state $\ket{G}$ at node $i$. Let
\begin{align}
    \ket{\pm}=\frac{1}{\sqrt{2}}(\ket{0}\pm\ket{1})
\end{align}
and
\begin{align}
    \ket{\pm i}=\frac{1}{\sqrt{2}}(\ket{0}\pm i\ket{1}).
\end{align}
\subsection{X Basis Measurement}
Let
\begin{align}
    O=H X.
\end{align}
Let $b_0\in N_a$ be any neighboring node of $a$. If $N_a$ is empty then $a$ is in the state $\ket{+}$ and nothing happens after measurement. The possible states after measurement when $N_a$ is not empty are 
\begin{align}
    &\op{+}:\ket{+}_a\otimes O_{b_0}Z_{N_{a}\setminus (N_{b_0}\cup b_0)}\ket{\tau_{b_0}(\tau_a\circ\tau_{b_0}(G)-a)}\\
    &\op{-}:\ket{-}_a\otimes O_{b_0}^\dagger Z_{N_{b_0}\backslash (N_{a}\cup a)}\ket{\tau_{b_0}(\tau_a\circ\tau_{b_0}(G)-a)}.
\end{align}
% For outcome $\op{+}_2$, we have that output state is proportional to
% \begin{align}
%     &\op{+}_2((I\otimes\op{0}\otimes I+ Z\otimes \op{1}\otimes Z)_{123})\ket{+++}_{123}\\
%     \notag&\rightarrow\\
%     &\frac{1}{\sqrt{2}}(Z\otimes Z+I\otimes I)\ket{++}_{13}\\
%     &=\frac{1}{\sqrt{2}}(\op{0}\otimes(Z+I)+\op{1}\otimes(I-Z))\ket{++}_{13}\\
%     &=\frac{2}{\sqrt{2}}(\op{00}+\op{11})\ket{++}_{13}\\
%     &=\ket{\Phi^+}.
% \end{align}
% $\ket{\Phi^+}$ can be transformed to graph state with $H_{1|2}$ and these are the local corrections.

% For outcome $\op{-}_2$, we have that the output state is proportional to
% \begin{align}
%     &\op{-}_2((I\otimes\op{0}\otimes I+ Z\otimes \op{1}\otimes Z)_{123})\ket{+++}_{123}\\
%     \notag&\rightarrow\\
%     &\frac{1}{\sqrt{2}}(-Z\otimes Z+I\otimes I)\ket{++}_{13}\\
%     &=\frac{1}{\sqrt{2}}(\op{0}\otimes(I-Z)+\op{1}\otimes(I+Z))\ket{++}_{13}\\
%     &=\frac{2}{\sqrt{2}}(\op{01}+\op{10})\ket{++}_{13}\\
%     &=\frac{1}{\sqrt{2}}(\ket{01}+\ket{10})=\ket{\Psi^+}.
% \end{align}
% Thus, the local corrections are $H_{1|2}X_{1|2}$ or $H_{2|1}X_{1|2}$.
\subsection{Y Basis Measurement}
Let
\begin{align}
    S=\begin{pmatrix}
        1 & 0\\
        0 & i
    \end{pmatrix}.
\end{align}
The possible states after a  Y basis measurement are 
\begin{align}
    &\op{i}_a:\ket{i}_a\otimes S_{N_a}\ket{\tau_a(G)-a}\\
    &\op{-i}_a:\ket{-i}_a\otimes S_{N_a}^\dagger\ket{\tau_a(G)-a}
\end{align}
up to a global phase.
\subsection{Z Basis Measurement}
The possible states after a  Z basis measurement are
\begin{align}
    &\op{0}_a:\ket{0}_a\otimes \ket{G-a}\\
    &\op{1}_a:\ket{1}_a\otimes Z_{N_a}\ket{G-a}.
\end{align}
% For outcome $\op{i}_2$, we have that the output state is proportional to
% \begin{align}
%     &\op{i}_2((I\otimes\op{0}\otimes I+ Z\otimes \op{1}\otimes Z)_{123})\ket{+++}_{123}\\
%     \notag&\rightarrow\\
%     &\frac{1}{\sqrt{2}}(-i Z\otimes Z+I\otimes I)\ket{++}_{13}\\
%     &=\frac{1}{\sqrt{2}}(\op{0}\otimes(I-iZ)+\op{1}\otimes(I+iZ))\ket{++}_{13}\\
%     &\notag=\frac{1}{\sqrt{2}}(\op{00}(1-i)+\op{01}(1+i)+\op{10}(1+i)\\
%     &\quad+\op{11}(1-i))\ket{++}_{13}\\
%     &\notag=\frac{1}{2\sqrt{2}}(\ket{00}(1-i)+\ket{01}(1+i)+\ket{10}(1+i)\\
%     &\quad+\ket{11}(1-i))\\
%     &\notag=\frac{1-i}{2\sqrt{2}}(\ket{00}+i\ket{01}+i\ket{10}+\ket{11}).
% \end{align}
% Thus, the local corrections are $S^\dagger_{1}S^\dagger_{2}$, where
% \begin{align}
%     S=\begin{pmatrix}
%         1 & 0\\
%         0 & i
%     \end{pmatrix}.
% \end{align}

% For outcome $\op{-i}_2$, we have that the output state is proportional to
% \begin{align}
%     &\op{-i}_2((I\otimes\op{0}\otimes I+ Z\otimes \op{1}\otimes Z)_{123})\ket{+++}_{123}\\
%     \notag&\rightarrow\\
%     &\frac{1}{\sqrt{2}}(i Z\otimes Z+I\otimes I)\ket{++}_{13}\\
%     &=\frac{1}{\sqrt{2}}(\op{0}\otimes(I+iZ)+\op{1}\otimes(I-iZ))\ket{++}_{13}\\
%     &\notag=\frac{1}{\sqrt{2}}(\op{00}(1+i)+\op{01}(1-i)+\op{10}(1-i)\\
%     &\quad+\op{11}(1+i))\ket{++}_{13}\\
%     &\notag=\frac{1}{2\sqrt{2}}(\ket{00}(1+i)+\ket{01}(1-i)+\ket{10}(1-i)\\
%     &\quad+\ket{11}(1+i))\\
%     &\notag=\frac{1+i}{2\sqrt{2}}(\ket{00}-i\ket{01}-i\ket{10}+\ket{11}).
% \end{align}
% Thus, the local corrections are $S_{1}S_{2}$.

\section{PCS X Fidelity and Postselection Rate}\label{appendix:PCS fidelity_postselect}
To simplify the following calculations, we rewrite the depolarizing channels as
\begin{align}
    \mathcal{E}(\rho)=\sum_{E_i}E_i\rho E_i^\dagger,
\end{align}
where $p$ is the probability of an error, $E_1=\sqrt{1-p}I$ and $E_2,E_3,$ and $E_4$ are $\sqrt{\frac{p}{3}}X, \sqrt{\frac{p}{3}}Y$, and $\sqrt{\frac{p}{3}}Z$, respectively. Note that this is a completely depolarizing channel when $p=\frac{3}{4}$. Let $\mathcal{E}_1(\rho)=\sum_iE_i\rho E_i^\dagger$ and $\mathcal{E}_2(\rho)=\sum_j G_j\rho G_j^\dagger$ be single qubit depolarizing channels with error probabilities $p_1$ and $p_2$, respectively.
\begin{widetext}
Define
\begin{align}
    N_{i,j}\equiv (\op{0}\otimes I+\op{1}\otimes P)^{(1,2)}(E_i\otimes E_j)^{(1,2)}(\op{0}\otimes I+\op{1}\otimes P)^{(1,2)},
\end{align}
\end{widetext}
where the superscripts denote the subspace that the operator acts on and $P$ is a Pauli matrix. Let us consider half of the full state, i.e., half of the Bell state with its ancilla. The state after the right PCS check and before the ancilla measurement is given by
\begin{align}
   \sum_{i,j}N_{i,j} (\op{+}\otimes\rho)N_{i,j}^\dagger,
\end{align}
where the first system is the ancilla and $\rho$ is half of the Bell state. Measuring the ancilla in the X basis and postselecting on the $\op{+}$ outcome we have
\begin{align}\label{eq:postselected_st1}
   \rho'=\frac{\sum_{i,j}\op{+}^{(1)}N_{i,j} (\op{+}\otimes\rho)N_{i,j}^\dagger\op{+}^{(1)}}{\tr(\sum_{i,j}\op{+}^{(1)}N_{i,j} (\op{+}\otimes\rho)N_{i,j}^\dagger)}.
\end{align}
Expanding, we have
\begin{widetext}
\begin{align}
    \op{+}^{(1)}N_{i,j}\op{+}^{(1)}&=\op{+}^{(1)}(\op{0}E_i\op{0}\otimes E_j+\op{0}E_i\op{1}\otimes E_jP+\op{1}E_i\op{0}\otimes PE_j\\
    &\notag+\op{1}E_i\op{1}\otimes PE_jP)\op{+}^{(1)}\\
    &=\frac{1}{2}\op{+}\otimes(\bra{0}E_i\ket{0} E_j+\bra{0}E_i\ket{1} E_jP +\bra{1}E_i\ket{0} PE_j+\bra{1}E_i\ket{1} PE_jP).
\end{align}
\end{widetext}
We can evaluate these for the different Pauli matrices. We have
% \begin{widetext}
\begin{align}
    \notag&\op{+}^{(1)}N_{k,j}\op{+}^{(1)}=\\
    &k=1,j: \frac{1}{2}\sqrt{1-p_1}\op{+}\otimes( E_j+PE_jP)\\
    &k=1,j=1: (1-p_1)\op{+}\otimes(I)\\
    &k=1,j=2: \frac{1}{2}\sqrt{1-p_1}\sqrt{\frac{p_1}{3}}\op{+}\otimes( X+PXP)\\
    &k=1,j=3: \frac{1}{2}\sqrt{1-p_1}\sqrt{\frac{p_1}{3}}\op{+}\otimes( Y+PYP)\\
    &k=1,j=4: \frac{1}{2}\sqrt{1-p_1}\sqrt{\frac{p_1}{3}}\op{+}\otimes( Z+PZP)\\
    %%%%%%%%%%%%%%%%%%%%%%%%%%%%%%%%%%%%%%%%%%%%%%%
    &k=2,j: \frac{1}{2}\sqrt{\frac{p_1}{3}}\op{+}\otimes( E_jP + PE_j)\\
    &k=2,j=1: \sqrt{\frac{p_1}{3}}\sqrt{1-p_1}\op{+}\otimes(P)\\
    &k=2,j=2: \frac{1}{2}\frac{p_1}{3}\op{+}\otimes( XP + PX)\\
    &k=2,j=3: \frac{1}{2}\frac{p_1}{3}\op{+}\otimes( YP + PY)\\
    &k=2,j=4: \frac{1}{2}\frac{p_1}{3}\op{+}\otimes( ZP + PZ)\\
    %%%%%%%%%%%%%%%%%%%%%%%%%%%%%%%%%%%%%%%%%%%%%%%
    &k=3,j: \frac{1}{2}\sqrt{\frac{p_1}{3}}i\op{+}\otimes( -E_jP + PE_j)\\
    &k=3,j=1: 0\\
    &k=3,j=2: \frac{1}{2}\frac{p_1}{3}i\op{+}\otimes( -XP + PX)\\
    &k=3,j=3: \frac{1}{2}\frac{p_1}{3}i\op{+}\otimes( -YP + PY)\\
    &k=3,j=4: \frac{1}{2}\frac{p_1}{3}i\op{+}\otimes( -ZP + PZ)\\
    %%%%%%%%%%%%%%%%%%%%%%%%%%%%%%%%%%%%%%%%%%%%%%%
    &k=4,j: \frac{1}{2}\sqrt{\frac{p_1}{3}}\op{+}\otimes( E_j - PE_jP)\\
    &k=4,j=1: 0\\
    &k=4,j=2: \frac{1}{2}\frac{p_1}{3}\op{+}\otimes( X - PXP)\\
    &k=4,j=3: \frac{1}{2}\frac{p_1}{3}\op{+}\otimes( Y - PYP)\\
    &k=4,j=4: \frac{1}{2}\frac{p_1}{3}\op{+}\otimes( Z - PZP).
\end{align}
The equations simplify when we define a specific Pauli check. Let $P=X$. Then we have
\begin{align}
    \notag&\op{+}^{(1)}N_{k,j}\op{+}^{(1)}=\\
    % &k=1,j: \frac{1}{2}\sqrt{1-p_1}\op{+}\otimes( G_j+PG_jP)\\
    &k=1,j=1: (1-p_1)\op{+}\otimes(I)\\
    &k=1,j=2: \sqrt{1-p_1}\sqrt{\frac{p_1}{3}}\op{+}\otimes( X)\\
    &k=1,j=3: 0\\
    &k=1,j=4: 0\\
    %%%%%%%%%%%%%%%%%%%%%%%%%%%%%%%%%%%%%%%%%%%%%%%
    % &k=2,j: \frac{1}{2}\sqrt{\frac{p_1}{3}}\op{+}\otimes( G_jP + PG_j)\\
    &k=2,j=1: \sqrt{\frac{p_1}{3}}\sqrt{1-p_1}\op{+}\otimes(X)\\
    &k=2,j=2: \frac{p_1}{3}\op{+}\otimes(I)\\
    &k=2,j=3: 0\\
    &k=2,j=4: 0\\
    %%%%%%%%%%%%%%%%%%%%%%%%%%%%%%%%%%%%%%%%%%%%%%%
    % &k=3,j: \frac{1}{2}\sqrt{\frac{p_1}{3}}i\op{+}\otimes( -G_jP + PG_j)\\
    &k=3,j=1: 0\\
    &k=3,j=2: 0\\
    &k=3,j=3: \frac{p_1}{3}i\op{+}\otimes(iZ)\\
    &k=3,j=4: \frac{p_1}{3}i\op{+}\otimes( -iY)\\
    %%%%%%%%%%%%%%%%%%%%%%%%%%%%%%%%%%%%%%%%%%%%%%%
    % &k=4,j: \frac{1}{2}\sqrt{\frac{p_1}{3}}\op{+}\otimes( G_j - PG_jP)\\
    &k=4,j=1: 0\\
    &k=4,j=2: 0\\
    &k=4,j=3: \frac{p_1}{3}\op{+}\otimes (Y)\\
    &k=4,j=4: \frac{p_1}{3}\op{+}\otimes (Z).
\end{align}
The $\op{+}^{(1)}N_{k,j}\op{+}^{(1)}$ terms define the effective error Kraus map. We can combine like terms. Let
\begin{align}\label{eq:eff_error_ops1}
    &\tilde E_1=\sqrt{(1-p_1)^2+\frac{p_1^2}{9}}I\\
    &\tilde E_2=\sqrt{(1-p_1)\frac{2p_1}{3}}X\\
    &\tilde E_3=\frac{\sqrt{2}p_1}{3}Y\\
    \label{eq:eff_error_ops4}&\tilde E_4=\frac{\sqrt{2}p_1}{3}Z.
\end{align}
% \end{widetext}
Substituting Eqs.~\eqref{eq:eff_error_ops1}-\eqref{eq:eff_error_ops4} into Eq.~\eqref{eq:postselected_st1} we get
\begin{align}
    \rho'=\frac{\sum_i \tilde E_i\rho \tilde E_i^\dagger}{\tr(\sum_i\tilde E_i\rho\tilde E_i^\dagger)}.
\end{align}
Substituting $p_1\rightarrow \frac{3p_1}{4}$ (to make $p_1=1$ a full depolarizing channel) and evaluating the trace in the denominator, we get a postselection rate of
\begin{align}
    c_1(p_1)=\frac{1}{2}(2+p_1(-2+p_1)).
\end{align}
Then we can describe the postselected state in terms of an effective error channel
\begin{align}
    \mathcal{E}'_2(\rho)=\rho'=\sum_iE'_i\rho E'^\dagger_i,
\end{align}
where $E'_i=\frac{\tilde E_i}{c_1}$ for all $i$. The fidelity of the postselected state is
\begin{align}
    F'_1=\frac{8+p_1(-12+5p_1)}{8+4p_1(-2+p_1))}.
\end{align}
This calculation only gives the output for the PCS process on one of the Bell state qubits. We can get these equations for the case where we apply PCS X checks on both qubits (two ancillas). Let the single qubit depolarizing channels on the other half of the state have error probability $p_2$. Since these two processes are independent, we can simply multiply $c_1(p_1)$ with $c_1(p_2)$ to get the overall postselection rate.
Thus,
\begin{align}\label{eq:postselection_pcs1}
    c_{12}=&c_1(p_1)c_2(p_2)\\
    \notag&=\frac{1}{4}[2+p_1(-2+p_1)][2+p_2(-2+p_2)].
\end{align}
For the fidelity, we need to be careful and realize that the $XX, YY,$ and $ZZ$ components of the combined channels contribute to the fidelity. Applying the effective error channel to both the qubits of the Bell state and taking the fidelity with respect to the Bell state, we get
\begin{align}\label{eq:fidelity_pcs1}
    F_{12}'=\frac{\splitfrac{[9 (p_1-2) p_1+10] p_2^2+2 (20-9 p_1) p_1 p_2}{+2 p_1 (5 p_1-12)-24 p_2+16}}{4 [(p_1-2) p_1+2] [(p_2-2)
   p_2+2]}.
\end{align}
These equations simplify when $p_1=p_2$. We get
\begin{align}
    c_{12}=\frac{1}{4}[2+(-2+p)p]^2
\end{align}
and
\begin{align}\label{eq:fidelity_both_PCSx_equalps1}
    F'_{12}=\frac{[4+3(-2+p)p]^2}{4[2+(-2+p)p]^2}.
\end{align}
Finally, the initial fidelity of the noisy Bell state (without PCS) and after local depolarization channels on both qubits is
\begin{align}\label{eq:initialF_nopcs1}
    F=1+\frac{3}{4}(-2+p)p.
\end{align}
Solving for $p$ in Eq.~\eqref{eq:initialF_nopcs1} we get
\begin{align}
    p=\frac{1}{3}(3-\sqrt{3}\sqrt{-1+4F}).
\end{align}
Substituting for $p$ yields
\begin{align}
    c_{12}=\frac{1}{9}(1+2F)^2
\end{align}
and
\begin{align}
    F'_{12}=\frac{9F^2}{(1+2F)^2}.
\end{align}
\vfill

\small

\framebox{\parbox{\linewidth}{
The submitted manuscript has been created by UChicago Argonne, LLC, Operator of 
Argonne National Laboratory (``Argonne''). Argonne, a U.S.\ Department of 
Energy Office of Science laboratory, is operated under Contract No.\ 
DE-AC02-06CH11357. 
The U.S.\ Government retains for itself, and others acting on its behalf, a 
paid-up nonexclusive, irrevocable worldwide license in said article to 
reproduce, prepare derivative works, distribute copies to the public, and 
perform publicly and display publicly, by or on behalf of the Government.  The 
Department of Energy will provide public access to these results of federally 
sponsored research in accordance with the DOE Public Access Plan. 
http://energy.gov/downloads/doe-public-access-plan.}}
%\bibliographystyle{unsrt}
%\bibliography{QEM}

\end{document}